\documentclass[12pt,preprint]{aastex}
\usepackage{graphicx}
\usepackage{amssymb}
\usepackage{amsmath}
\usepackage{natbib}
\usepackage{hyperref}

\shortauthors{Durant et al.}

\newcommand{\be}{\begin{equation}}
\newcommand{\ee}{\end{equation}}

\begin{document}

\title{Hubble Space Telescope detection of the double pulsar system J0737$-$3039 in the far-ultraviolet}

\author{Martin Durant\altaffilmark{1}, Oleg Kargaltsev\altaffilmark{2}, 
 and George G.\ Pavlov\altaffilmark{3}}

\altaffiltext{1}{Department of Medical Biophysics, Sunnybrook Hospital M6 623, 2075 Bayview Ave., Toronto M4N 3M5, Canada; mdurant@sri.utoronto.ca}
\altaffiltext{2}{Department of Physics, The George Washington University, 725 21st St NW, Washington, DC 20052,   USA; kargaltsev@email.gwu.edu}
\altaffiltext{3}{Department of Astronomy and Astrophysics, Pennsylvania State University,
University Park, PA 16802, USA; pavlov@astro.psu.edu}


\begin{abstract}
We report on detection of the double pulsar system J0737$-$3039 in the far-UV with the ACS/SBC detector aboard {\sl HST}. 
We measured the energy flux $F = (4.5\pm 1.0)\times 10^{-17}$ erg cm$^{-2}$ s$^{-1}$ in the 1250--1550 \AA\ band, which corresponds to the extinction-corrected luminosity $L\approx 1.5\times 10^{28}$ erg s$^{-1}$ for the distance $d=1.1$ kpc and a plausible reddening $E(B-V)=0.1$. 
If the detected emission comes from the entire surface of one of the neutron stars with a 13 km  radius, the surface blackbody temperature is in the range $T\simeq(2$--$5)\times 10^5$ K for a reasonable range of interstellar extinction.
Such a temperature requires an internal heating mechanism to operate in old neutron stars, or it might be explained by heating of the surface of the less energetic Pulsar B by the relativistic wind of Pulsar A.
If the far-UV emission is non-thermal (e.g., produced in the magnetosphere of Pulsar A), its spectrum exhibits a break between the UV and X-rays. 
\end{abstract}
\maketitle

\section{Introduction}
Double Neutron Star Binaries (DNSBs) are very rare objects (currently, 8 DNSBs are known) thought to be formed in consequent supernova explosions in the course of evolution of a massive binary system (e.g., \citealt{2004PhRvL..93n1101S}). 
It is generally believed that once the first neutron star (NS) is formed and the second star evolves further, it starts to transfer matter onto the primary NS.
The accretion results in a spin-up of the NS, which becomes a millisecond (recycled) pulsar. At some point the second star explodes as a supernova and, if the explosion does not disrupt the binary or disintegrate the second star, a DNSB is formed.

J0737$-$3039 (J0737 hereafter) is the only known example of a DNSB system in which both NSs have been detected as pulsars \citep{2004Sci...303.1153L}.
Observations of DSNBs in general are of great importance since they allow one to accurately measure the masses of both NSs and test the predictions of General Relativity (GR) or other models of gravity. 
Having two precise clocks orbiting each other in a tight binary and the fortuitous viewing angle, nearly edge-on (inclination angle $i\simeq 89^{\circ}$) greatly improves the precision and expands the number of potentially measurable quantities (e.g., the NS moment of inertia and higher-order terms in GR); see \citet{2008ARA&A..46..541K} for a review. 
In addition to the radio timing, optical, UV and X-ray observations have the potential to provide additional useful constraints 
on the NS radius through measurements of thermal emission from the NS surface, particularly 
when the  distance to the NS is accurately measured through its parallax  (as it is for  J0737, $d=1.25_{-0.16}^{+0.22}$\,kpc, \citealt{2009Sci...323.1327D}; or $d=1.1_{-0.1}^{+0.2}$\,kpc after Lutz-Kelker correction, \citealt{2012ApJ...755...39V}). 
UV and X-ray observations can also measure  the NS surface temperature (e.g., \citealt{2002nsps.conf..273P}), which, together with the mass and radius information, can provide tight constraints on the unknown equation of state of matter in the superdense NS interior \citep{2004ARA&A..42..169Y}. 

Radio observations \citep{2003Natur.426..531B,2004Sci...303.1153L} have shown that the binary consists of the recycled
Pulsar A ($P_A=22.7$ ms, $\dot{E}_{A}=5.8\times 10^{33}$ erg s$^{-1}$, $\tau_{A}=P_{A}/2\dot{P}_{A}=210$\,Myr, $B_A=6.4\times10^9$\,G) and
the ordinary old Pulsar B ($P_B=2.8$ s, $\dot{E}_{B}=1.6\times 10^{30}$ erg s$^{-1}$, $\tau_{B}=P_{B}/2\dot{P}_{B}=50$\,Myr,
 $B_B=1.6\times10^{12}$\,G) locked in a tight 2.4-hour orbit with a maximum separation of only $3$
light-seconds and an eccentricity $e\approx0.09$.
The nearly edge-on geometry of J0737 allowed \citet{2004Sci...303.1153L} to detect a radio eclipse of Pulsar A by the magnetosphere of Pulsar B, likely resulting from synchrotron absorption in the magnetosphere of the latter \citep{2005ASPC..328...95A,2004MNRAS.353.1095L}.
Even more interesting was the finding that the pulsed radio emission of Pulsar B was seen only during two short orbital phase intervals  before and after Pulsar A's superior conjunction. 
These changes strongly suggest that the magnetosphere (and possibly atmosphere) of Pulsar B is influenced by the wind or radiation of the more powerful Pulsar A \citep{2004Natur.428..919J,2004ApJ...615L.137D,2004ApJ...614L..53Z}.
Pulsar A dominates the energetic output of the system (by over a factor of 3000 in spin-down power), and so it is likely to dominate the non-thermal output as well\footnote{The emission, however, may come not only from Pulsar A's magnetosphere, but also from an intrabinary shock or Pulsar B's magnetosphere, activated by Pulsar A's wind.}.

J0737 is also the first of the two known DNSBs detected in X-rays (the second being B1534+12; \citealt{2006ApJ...646.1139K,2011ApJ...741...65D}). 
Following observations with {\sl Chandra} ACIS (10 ks, \citealt{2004ApJ...605L..41M}, and 80 ks, \citealt{2008ApJ...680..654P}) and {\sl XMM-Newton} (50 ks; \citealt{2004ApJ...612L..49P,2004ApJ...613L..53C}), a 230-ks {\sl XMM-Newton} observation was undertaken 
(\citealt{2008ApJ...679..664P}, P+08 hereafter).  
The derived spectrum allows for several two-component or three-component models, e.g., two blackbodies (BB+BB), blackbody plus power-law (BB+PL), or BB+BB+PL. 
The reason for this is the relatively narrow energy range of the X-ray data (0.3--8\,keV, with best sensitivity in 0.5--2\,keV) and the large number of correlated model parameters.

\cite{2013ApJ...768..169G} presented the detection of J0737 in GeV $\gamma$-rays with {\sl Fermi} LAT. Pulsations from Pulsar A were detected, but not from Pulsar B. 
A double-peaked pulsation profile was found, with a $\gamma$-ray efficiency in the typical range for millisecond pulsars.
Although significantly detected, the inferred spectrum was highly uncertain, with photon index $\Gamma<1.3$ and cut-off energy $E_c=0.4\pm0.4$\,GeV.
The magnetic inclination angle, $\alpha$, and viewing angle, $\zeta$, were both found to be near 90$\degr$, consistent with the radio modeling. 

There was only one  deep  observation  J0737 in the optical (\citealt{2012ApJ...749...84F}; F+12 hereafter), employing the high resolution camera (HRC) of the Advanced Camera for Surveys (ACS) aboard the {\sl Hubble Space Telescope} ({\sl HST}). F+12 used a coronograph to obstruct a nearby  ($\approx3\arcsec$ away)  bright ($V=13$, late F-type) star, but the target was not detected.

To reduce the relative brightness of the F-star, we observed J0737 in the far ultraviolet (FUV).
In this Letter we describe the observation,  present the discovery of the FUV emission, and discuss some implications.

\section{Observations and Results}
	
J0737 was observed with HST on 2012 December 12 in the FUV using the Solar-blind Channel (SBC) of ACS  (34\farcs6$\times30\farcs8$ field of view, 0\farcs034$\times$0\farcs030 pixel scale), using long-pass filter F125LP with peak transmission around 1300\,\AA\ and sharp cut-off to shorter wavelengths\footnote{See {\tt http://www.stsci.edu/hst/acs/documents/handbooks/cycle20/c10\_ImagingReference40.html}}. 
A total exposure time of 5553\,s was acquired across three orbits in the Earth's shadow, where the geo-coronal FUV background is greatly reduced. 
The resulting combined 6\arcsec$\times$6\arcsec\ image around the position of J0737 is shown in Figure \ref{fig:field}. 
In addition to the bright field F-star, we detected a  faint point  source, which we show below to be the FUV counterpart to J0737.

We also filled the remainder of the orbits with imaging using filter F140LP\footnote{See {\tt http://www.stsci.edu/hst/acs/documents/handbooks/cycle20/c10\_ImagingReference41.html}}, for a total exposure time of 1845\,s, because this filter cuts off the geocoronal background (including the oxygen lines) seen at shorter wavelengths when not in the Earth's shadow.

\subsection{Astrometry}

\begin{figure}
\centering
\includegraphics[width=0.7\linewidth]{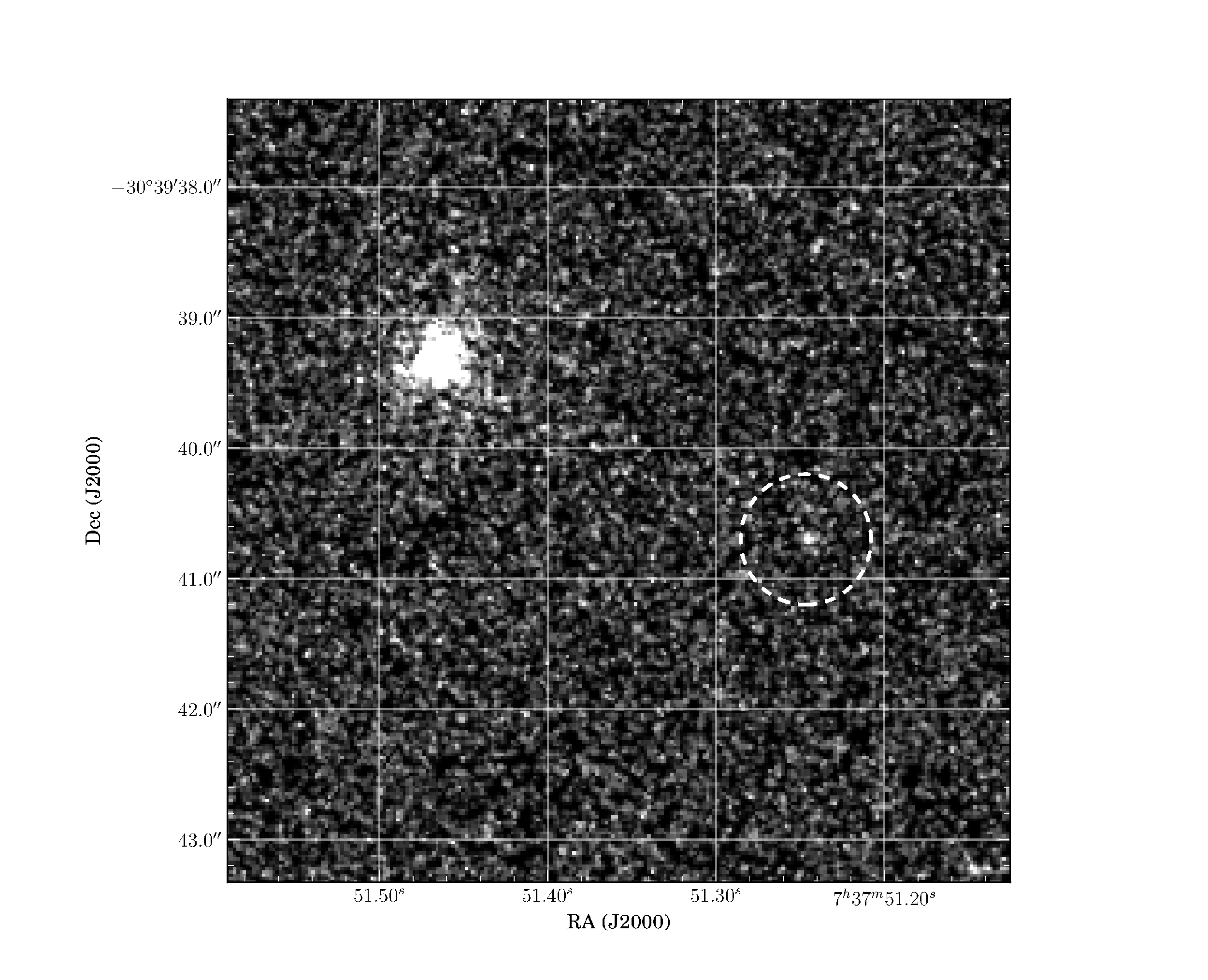}
\includegraphics[width=0.7\linewidth]{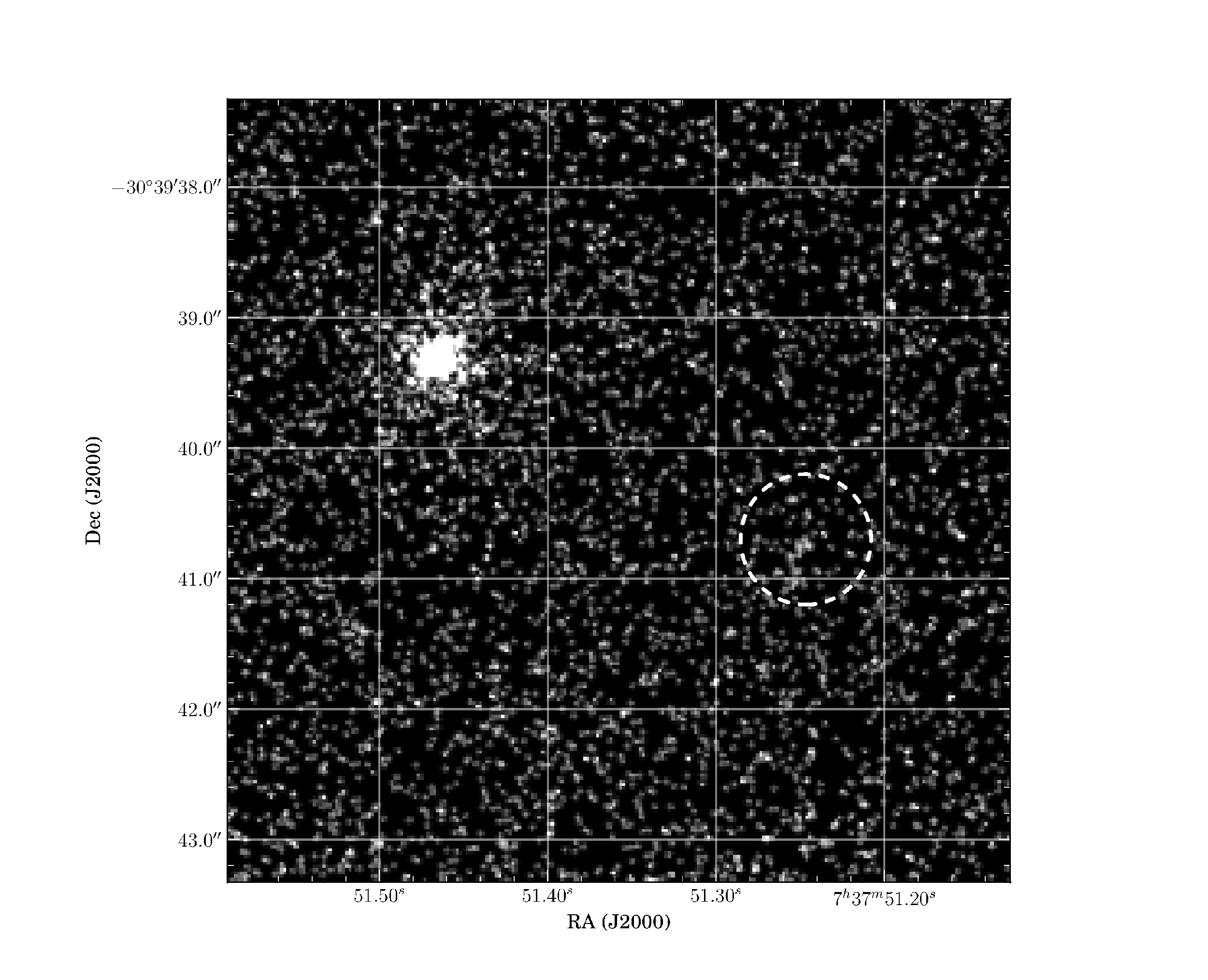}
\caption{Image of the field of J0737 in F125LP (top) and F140LP (bottom). 
The world coordinates of the field have been adjusted such that the field F-star aligns with projected UCAC4 catalog position (see text).
The dashed circle (radius of 0\farcs5) is centered on the predicted radio position of J0737.
}
\label{fig:field}
\end{figure}

We based our astrometry on the F-star position. The most accurate astrometric information for this star is given in the UCAC4 catalog \citep{2012yCat.1322....0Z}. 
The catalog position of the star 297-026440 is $\alpha=114\fdg464\,512\,4(30)$, $\delta=-30\fdg660\,996\,7(47)$ at the 
epoch of 2000.0, its proper motion is $\mu_\alpha=-18.3\pm1.7$\,mas\,yr$^{-1}$, $\mu_\delta=20.1\pm1.7$\,mas\,yr$^{-1}$.
Calculating the star's position at the epoch of the {\sl HST} observations (2012.95), we find that the {\sl HST} coordinates are offset with respect to the UCAC4 coordinates by 0\farcs29$\pm$0\farcs03 in R.A. and 1\farcs33$\pm$0\farcs03 in Dec. 
We, therefore, adjusted the WCS of the image, by simple translation. 
The pulsar's position also needs to be corrected for proper motion; from the information in \citet{2009Sci...323.1327D}
we found $\alpha=114\fdg463\,527\,7(12)$, $\delta=-30\fdg661\,305\,8(4)$ at the epoch of the {\sl HST} observations.

In the corrected F125LP image we found a source located almost exactly at the predicted position of J0737 (Figure \ref{fig:field}, top). 
The difference between the star-pulsar offset as predicted (p) by the calculations above and as measured (m) from the image is $(\alpha_\ast^p - \alpha_{\rm PSR}^p)\cos\delta - (\alpha_\ast^m - \alpha_{\rm PSR}^m)\cos\delta = 2\farcs813 - 2\farcs839 = 0\farcs026$, $(\delta_\ast^p - \delta_{\rm PSR}^p) - (\delta_\ast^m - \delta_{\rm PSR}^m) = 1\farcs374 - 1\farcs370 = 0\farcs004$.
The offset is smaller than the uncertainty of the predicted J0737 position, dominated by the error in the reference star's proper motion over $\sim$13\,yr, about 0\farcs03 in each coordinate (one-sigma), under one image pixel.
This source is not present in any optical or IR catalog, nor in the deep optical {\sl HST}  images of  F+12.
With the low surface density of FUV sources in the field, a chance coincidence would be extremely unlikely and, hence, we conclude that the FUV source we discovered is indeed J0737.

In the F140LP image (also corrected using UCAC4; see Figure 1, bottom), a faint enhancement in the count rate is seen at a very similar position. The centroid is poorly determined, but about 0\farcs07 distant from the F125LP counterpart of J0737. As this could be due to a noise fluctuation, we consider this as a tentative detection.

\subsection{Photometry and Spectrum}\label{sec:spec}

\begin{figure}
\centering
\includegraphics[width=0.7\linewidth]{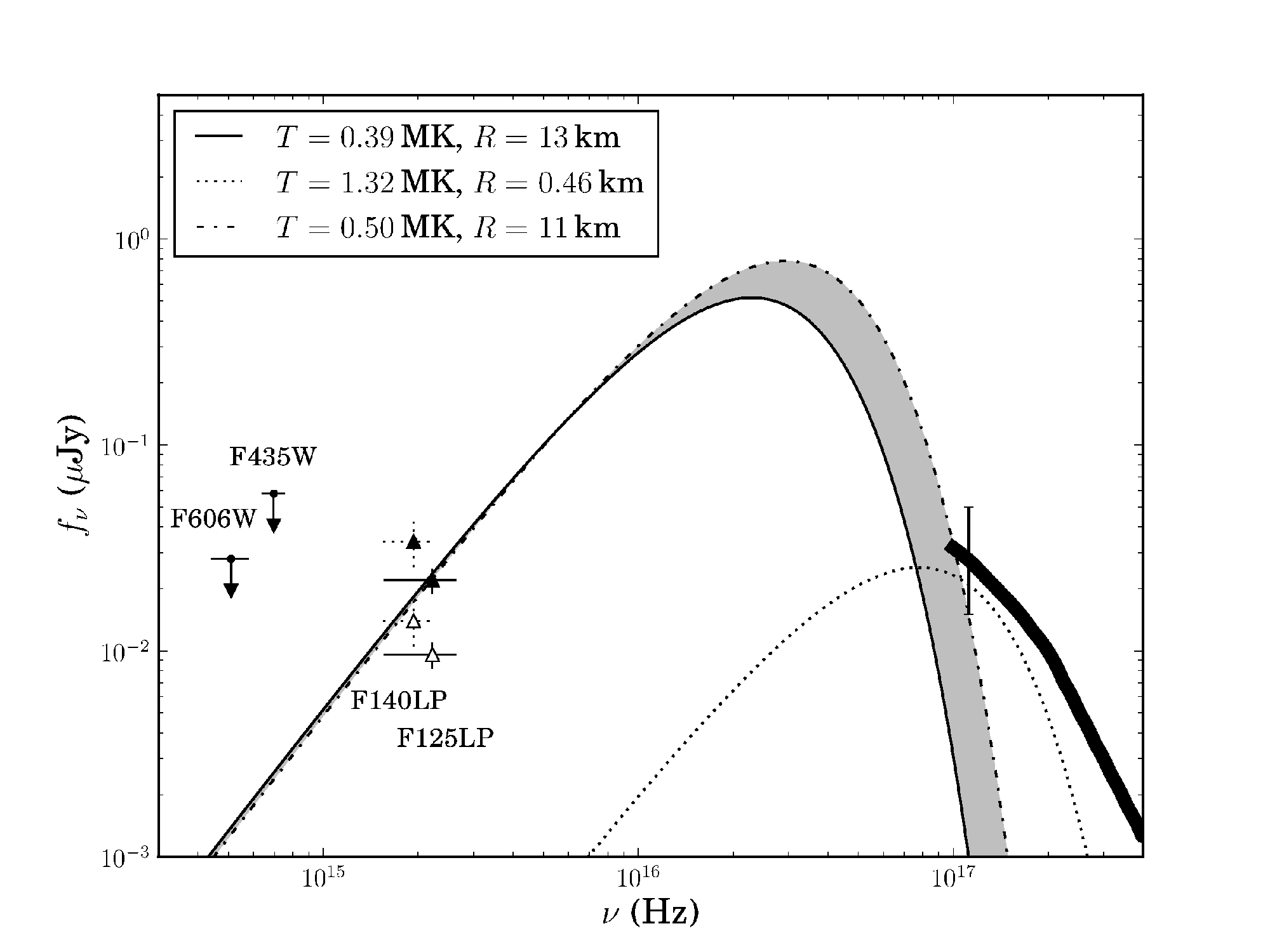}
\caption{
Measurements and models for the spectral flux of J0737. 
The open and filled triangles show the measured and extinction-corrected FUV fluxes, respectively, in the F125LP and F140LP filters (solid and dotted error-bars, respectively) for $E(B-V)=0.1$.
The upper limits in two optical filters are from F+12.
The unabsorbed X-ray spectrum (P+08) is shown by the thick line; the dotted line shows the cold BB component from the BB+BB model for the phase-integrated X-ray spectrum.
The solid and dashed lines show examples of BB models which best fit the F125LP detection for $E(B-V)=0.1$. The radii in the legend are for $d=1.1$ kpc.}
\label{fig:spec}
\end{figure}

In an aperture of radius 0\farcs0875, we measure the F125LP count rate of 0.0084\,cts\,s$^{-1}$ (corresponding to about 47 counts in total
\footnote{The total number of counts in the aperture is not an integer due to the drizzling procedure with which the three images were combined into the final image.}), 
of which 0.0032\,cts\,s$^{-1}$ are from the background (measured in a 50\arcsec\ radius circle and scaled to the 0\farcs0875 aperture). 
The detection significance is 4.2$\sigma$. 
The aperture correction was measured from the bright field F-star, to correct to the 0\farcs5 radius aperture for which  the photometry calibration baseline is established. 
We found a correction factor of 1.52
\footnote{The ACS instrument handbook recommends measuring the correction  rather than using tabulated values, see {\tt http://www.stsci.edu/hst/acs/documents/handbooks/current/c05\_imaging7.html}}.
The final derived source count rate is 0.0079$\pm$0.0018\,cts\,s$^{-1}$.
Using the same method, the derived count rate for the possible faint F140LP source (using the nominal aperture center, without centroiding) is 0.005$\pm$0.002\,cts\,s$^{-1}$, at 2.5$\sigma$ significance.

We used the spectral modeling package {\tt pysynphot} to calculate the count-rate expected in F125LP for BB model spectra and a range of temperatures and extinctions. 
The {\tt pysynphot} package integrates the model spectrum over the asymmetric instrumental band-pass (sensitivity function) and returns the count rate for given model parameters.
Comparing the simulated count rates with the measured one, we derive a best-fit temperature for each assumed value of extinction and emission area.
We used {\tt pysynphot}'s standard Galactic extinction from \citet{1989ApJ...345..245C}.
For apparent emitting area of a $R=13$\,km sphere, as would be expected from one of the pulsars, with a distance $d=1.1$\,kpc and plausible reddening $E(B-V)=0.1$, we find $T=(3.9\pm0.3)\times10^5$\,K from the F125LP detection.
With this assumed spectrum, the spectral flux is $f_\nu = 0.010\pm0.002$\,$\mu$Jy ($f_{\nu,{\rm corr}}=0.022\pm0.005$\,$\mu$Jy, corrected for extinction) at the effective wavelength $\lambda_{F125}=1414$\,\AA.
The integrated flux in the 1250--1550\,\AA\ band is $F_{\rm F125} 
= (4.6\pm 1.0)\times10^{-17}$\,erg\,s$^{-1}$\,cm$^{-2}$, or $F_{\rm F125,corr} 
= (1.1\pm 0.3)\times10^{-16}$\,erg\,s$^{-1}$\,cm$^{-2}$  after correcting for extinction (corresponding to a luminosity $L_{\rm F125}\approx 1.5\times 10^{28}$\,erg\,s$^{-1}$ at $d=1.1$\,kpc). 

For a PL that is flat (frequency-independent) in $F_\nu$ (one possible non-thermal spectrum), we find $f_{\nu,corr} = 0.021\pm0.005$\,$\mu$Jy, implying $F_{\rm F125}= (4.4\pm 1.0)\times10^{-17}$\,erg\,s$^{-1}$\,cm$^{-2}$, $F_{\rm F125,corr}= (1.1\pm 0.3)\times10^{-16}$\,erg\,s$^{-1}$\,cm$^{-2}$  (corresponding to $L_{\rm F125} \approx 1.5\times10^{28}$\,erg\,s$^{-1}$).

The derived spectral flux for the tentative  F140LP detection is $f_\nu = 0.014\pm0.006$\,$\mu$Jy ($f_{\nu, {\rm corr}}=0.0340\pm0.014$\,$\mu$Jy) at the effective wavelength $\lambda_{F140}=1507$\,\AA, for a thermal model.
In Figure  \ref{fig:spec} we also include the flux limits of  F+12 in the optical and the extinction-corrected X-ray spectrum from P+08 in the X-rays; the rough uncertainty at the low-energy end of the X-rays associated with extinction in the range $N_{\rm H}=(2-4)\times10^{20}$\,cm$^{-2}$ is indicated by an error-bar.
We also show the cool component of the BB+BB fit of the phase-integrated X-ray spectrum (P+08).

\section{Discussion}

The excellent positional coincidence of the source firmly detected in the F125LP  image leaves no doubt that it is the counterpart of J0737.  The detection in F140LP is marginal. 
Given the large uncertainties and overlapping bands, the estimated F140LP flux is consistent with the F125LP flux, but it does not add useful spectral information. If treated as an upper limit, it is less restrictive than those in the optical.
Therefore, we do not include it in further analysis.
 
The FUV emission can come from the surface of one (or  both) NSs, in which case it should have a thermal spectrum. 
Alternatively, the emission can be nonthermal, coming either from the magnetosphere of Pulsar A (the other pulsar is not energetic enough) or from a shock  powered by the interaction of the Pulsar A's 
wind with Pulsar B's magnetosphere (e.g., \citealt{2013arXiv1310.2204L}) or the circumbinary medium (for the bow-shock scenario, see, e.g., \citealt{2004ApJ...609L..17G,2004ApJ...614L..53Z}). 
With just one measurement, we cannot definitively discriminate between these possibilities. 
Below we discuss constraints on the nature of the FUV emission based on the existing multiwavelength data.

The measured FUV flux, as well as the optical upper limits (F+12), are clearly inconsistent with the extrapolation of the X-ray PL component ($\Gamma_X\approx 3$; P+08), which overpredicts the optical-FUV fluxes by several orders of magnitude.
Comparing the FUV flux with the optical limits, we find that the  optical-FUV spectrum cannot decrease with frequency faster than $F_\nu \propto \nu^{-0.2}$, for the plausible reddening $E(B-V) =0.1$ (i.e., the optical-FUV photon index $\Gamma_{\rm opt-FUV} < 1.2$). 
It means, in particular, that, if the FUV emission is nonthermal, there is a break in the spectral slope between the  FUV and X-ray bands, $|\Delta\Gamma| \gtrsim 2$. 
Such a break has been seen in spectra of young pulsars (e.g., the Vela pulsar; \citealt{2005ApJ...627..383R,2001ApJ...552L.129P}), but it remains unclear whether or not a similar break is present in old and recycled pulsars (e.g., \citealt{2004ApJ...616..452Z,2011ApJ...743...38D}).
Extrapolation of the {\sl Fermi} LAT spectrum \citep{2013ApJ...768..169G} toward the optical  falls below the measured FUV flux; however,  the $\gamma$-ray spectrum is very uncertain.

The FUV luminosity $L_{\rm F125}\approx1.5  \times10^{28}$ erg s$^{-1}$, for a flat spectrum and $E(B-V)=0.1$, corresponds to the radiative efficiency $\eta_{\rm F125} = L_{\rm F125}/\dot{E_{A}}\approx2.3 \times 10^{-6}$  (we use $\dot{E}_A$ in this estimate because Pulsar A is by far the  dominant source of power in the system), consistent with typical values for optical-UV efficiency (e.g., \citealt{2006AdSpR..37.1979Z}).
The FUV luminosity  is a factor of $\sim 700$ lower than the nonthermal  X-ray luminosity in 0.2--10 keV, which is also typical for pulsars detected in the optical-UV \citep{2004ApJ...616..452Z}.

\begin{figure}
\centering
\includegraphics[width=0.7\linewidth]{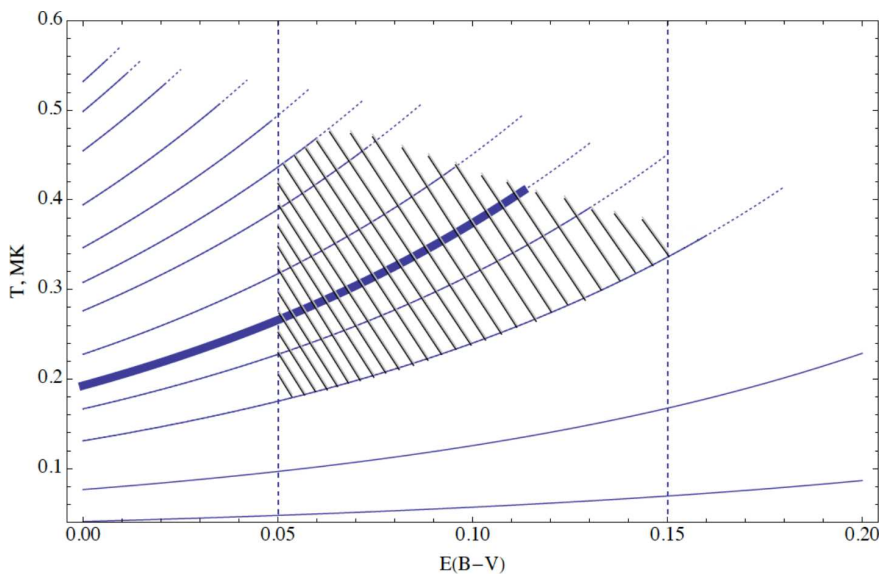}
\caption{Dependence of BB temperature on reddening for $R_{13}/d_{1.1}$=0.55, 0.57, 0.6, 0.65, 0.7, 0.75, 0.8, 0.9, 1.0, 1.1, 1.3, 2.0, and 4.0 (top to bottom; the thick curve corresponds to   $R_{13}/d_{1.1}$=1) for the thermal interpretation of the FUV emission. Each of the curves is drawn up to the $E(B-V)$ and $T$ values above which the extrapolated thermal FUV spectra start exceeding the measured X-ray spectrum (the dotted extensions for each line reflect measurement uncertainties).
The hatched area corresponds to the plausible parameter domain. }
\end{figure}

The FUV emission can also be interpreted as thermal emission from the NS surface (for simplicity, below we assume that it is dominated by one NS).
The unabsorbed FUV flux and the corresponding surface temperature $T$ depend on the reddening $E(B-V)$ and the wavelength dependence of extinction $A_\lambda$ (for instance, according to \citealt{1989ApJ...345..245C},  $A_\lambda = 8.7\,E(B-V)$  for the effective $\lambda=1414$ \AA\ of the F125LP filter).
The range of $N_H$ found from X-ray fits (P+08) suggests  $A_V\approx N_H/(1.8\times10^{21}~{\rm cm^{-2}})=0.17$--$0.43$ for  V-band extinction,  or $E(B-V)=A_V/R_V\approx 0.05$--0.15 for an average Galactic $R_V=3.1$.
For this range we obtain  $T =0.28$--$0.54$ MK  for the NS radius $R=13$\,km and $d=1.1$ kpc,
 which corresponds to bolometric luminosities $L_{\rm bol} = 4\pi R^2 \sigma T^4 = (0.7$--$10)\times 10^{31}$ erg s$^{-1}$.

For the average reddening of the above range, $E(B-V)=0.1$, we obtained  $T\approx0.4$ MK ($L_{\rm bol} \approx 3\times 10^{31}$ erg s$^{-1}$).
This temperature is similar to the cool component temperature, $T_c$, found by P+08 in a BB+BB+PL fit of the phase-resolved X-ray spectrum and ascribed to Pulsar B.
However, the normalization of that cool X-ray component implies an implausible emitting radius, $R_c\simeq33$ km, at $d= 1.1$ kpc.
This suggest that either the parallax measurement greatly overestimated the distance or, more likely, that the X-ray models were too simplistic to capture  the emission properties of this complex binary system.

The requirement that the extrapolation of the thermal FUV spectrum to higher energies should not exceed the measured X-ray spectrum puts additional constraints on the extinction, 
radius-to-distance ratio  $R_{13}/d_{1.1}$ (where $R_{13}=R/13$\,km and $d_{1.1}=d/1.1$\,kpc) and the NS surface temperature.  The region of allowed parameters is shown in Figure 3,  with the most plausible  values 
within the shaded area. 
The optical limits  
(F+12) constrain the parameters
(e.g., $T\gtrsim 17,000$--$24,000$ K, depending on $E(B-V)$) 
only at implausibly  large $R_{13}/d_{1.1}$ (not shown in Figure 3).

The inferred surface temperature is larger than the predictions of standard cooling models, $T<10^4$ K for pulsars older than 10 Myr (both pulsars in J0737 are believed to be significantly older; \citealt{2007MNRAS.379.1217L}).
This is also substantially higher than the temperature $T\approx0.15$ MK  of  the $6$\,Gyr-old millisecond PSR J0437--4715  \citep{2004ApJ...602..327K,2012ApJ...746....6D}.
Such surface temperatures require a heating mechanism to operate in the NS interior (see, e.g., \citealt{2010A&A...522A..16G}).
We defer further  discussion of NS heating/cooling models until the thermal nature of the FUV spectrum is confirmed.

For J0737, there may be an alternative heating mechanism (which is not relevant to isolated pulsars): the surface of Pulsar B could be heated by  Pulsar A's wind if  the ultra-relativistic particles of the wind  penetrate Pulsar B's magnetosphere and precipitate onto the surface.
The bolometric luminosity of Pulsar B should be  smaller than the fraction of Pulsar A's spin-down power intercepted by the Pulsar B's magnetosphere:
$L_{B,{\rm bol}} < \dot{E}_A (R_{\rm B,LC}/2d_{AB})^{2}\sim3\times10^{31}$\,erg\,s$^{-1}$ for an isotropic wind, where $R_{\rm LC,B}=1.3\times10^{10}$ cm is Pulsar B's light cylinder radius, and  $d_{AB}=9\times10^{10}$ cm is the average separation between the pulsars.
Since the radiative efficiency is typically a small fraction of the total power available, this estimate shows that  the observed emission is unlikely to come from the Pulsar B's surface heated by the Pulsar A's wind.
It is more likely that we see internal NS reheating operating in Pulsar A.

\section{Conclusion}

We discovered FUV emission from J0737 and measured the flux in two broad filters.
If the observed emission is thermal,  it would imply a pulsar surface temperature that is much higher than predicted by standard coiling curves.
Alternatively, the FUV emission could be non-thermal, but this would require a broken PL spectrum which decreases faster in X-rays than in the optical-UV with increasing  frequency.
Further  observations are needed to firmly establish the nature of the FUV emission from J0737.

\medskip\noindent{\bf Acknowledgments:}
This work was supported by National Aeronautics Space Administration through the HST-GO-12494.04-A award issued by the Space Telescope Science Institute.   

\bibliography{database}

\end{document}